\documentclass[lettersize,onecolumn]{IEEEtran}
\usepackage{amsmath,amsfonts,amsthm,amssymb}
\usepackage{algorithm}
\usepackage{hyperref}
\usepackage{algpseudocode}
\usepackage{array}
\usepackage[caption=false,font=normalsize,labelfont=sf,textfont=sf]{subfig}
\usepackage{textcomp}
\usepackage{url}
\usepackage{verbatim}
\usepackage{graphicx}
\usepackage{cite}
\usepackage{wrapfig}
\usepackage{dsfont}
\usepackage{tikz}
\usetikzlibrary{arrows.meta,calc,positioning}
\newtheorem{theorem}{Theorem}

\newtheorem{remark}{Remark}
\newtheorem{example}{Example}

\newif\ifCOM
\COMfalse

\begin{document}

\title{Structure of kissing arrangements in ${\mathbb R}^{12}$ and a place for the $841$st sphere}

\author{Rustem~Takhanov, Zhenisbek~Assylbekov and Stanislav~Yun
\thanks{R.~Takhanov and S.~Yun are with the Mathematics Department, Nazarbayev University, and Nazarbayev University Research Administration, Astana, Kazakhstan, Zh.~Assylbekov is with Purdue University Fort Wayne, e-mail: rustem.takhanov@nu.edu.kz.}
}



\maketitle
\begin{abstract}
Most currently known kissing arrangements of size $840$ in $\mathbb R^{12}$  share a common structure. They consist of $60$ vectors supported on $\mathbb R^6\times\{\mathbf 0\}$, another $60$ vectors supported on $\{\mathbf 0\}\times\mathbb R^6$, and $720$ additional \emph{bridge vectors}. The bridge vectors encode the interaction between the two six-dimensional factors and are constructed from the unique $1$-factorization of the complete graph $K_6$.

In this paper we investigate kissing arrangements of this type while keeping the bridge vectors fixed. We show that each $60$-point block admits substantial flexibility: $12$ of its vectors may be chosen as the signed coordinate vectors $\pm e_i$, while the remaining $48$ vectors may vary within a positive-dimensional family of configurations, which we call $48$-systems. As a consequence, we obtain infinitely many pairwise non-isometric kissing arrangements of size $840$ in $\mathbb R^{12}$.

The geometric freedom revealed by these constructions provides new insight into the local structure of extremal configurations. Exploiting this structure, we develop a specialized initialization scheme for logarithmic Riesz energy optimization. Starting from such structurally informed initial configurations, we numerically construct a kissing arrangement of size $841$ in $\mathbb R^{12}$.
\end{abstract}


\section{Introduction}

The \emph{kissing number problem} is one of the oldest and most fundamental questions in discrete geometry. For a positive integer $n$, the kissing number $\tau_n$ is defined as the maximum number of unit spheres in $\mathbb R^n$ that can simultaneously touch a fixed unit sphere without overlapping. Equivalently, $\tau_n$ is the largest size of a subset $X\subset S^{n-1}$ such that
\[
    x^\top y \le \frac12
    \qquad
    \text{for all distinct } x,y\in X.
\]
Such a set $X$ is called a \emph{kissing arrangement}. Determining the values of $\tau_n$ and understanding the structure of extremal kissing arrangements constitute central problems in sphere packing and coding theory.

The history of the problem can be traced back to the famous discussion between Newton and Gregory concerning the three-dimensional case. 
Despite centuries of study, exact kissing numbers are known only in a handful of dimensions. The values of $\tau_n$ have been determined for
$n=1,2,3,4,8,24$~\cite{Schütte1952,2003RuMaS,zbMATH03680654,ODLYZKO1979210},
with the dimensions $8$ and $24$ relying on the exceptional structures of the $E_8$ lattice and the Leech lattice. Outside these special cases, only upper and lower bounds are available.

Considerable effort has been devoted to determining kissing numbers in low dimensions. Considerable effort has been devoted to determining kissing numbers in low dimensions. In particular, dimensions up to $9$ have been the subject of extensive investigations, leading to increasingly strong lower bounds and numerous candidate extremal configurations~\cite{Korkine1873,Leech_Sloane_1971,Conway1995}. The long history and intensity of research in these dimensions make substantial future improvements appear less likely than in higher dimensions that have received comparatively less attention.

Interest in the kissing number problem has experienced a resurgence in recent years. New lower bounds have been established in dimensions
$n= 10,11,13,14,17,18$,  and $19$~\cite{GANZHINOV202512,zbMATH01467813,cohn2026improvedkissingnumbersseventeen,ho2026newlowerboundkissing},
demonstrating that progress remains possible even in dimensions that had resisted improvement for decades. Some of these advances have relied on large-scale computational searches and optimization techniques.

A particularly intriguing recent development is the use of artificial intelligence methods in the search for large kissing arrangements. For example, a new record in dimension $11$ was obtained by the AlphaEvolve~\cite{novikov2025alphaevolvecodingagentscientific} and the EinsteinArena~\cite{bianchi2026harnessingcollectiveintelligenceai} systems, illustrating the growing role of machine-assisted discovery in discrete geometry.

Against this background, dimension $12$ stands out as a remarkable exception. For reasons that are still poorly understood, the best known lower bound
\[
    \tau_{12}\ge 840
\]
has resisted improvement since its discovery in 1971~\cite{Leech_Sloane_1971}. No larger kissing arrangement has been found despite substantial advances in computational power, optimization methods, and our understanding of spherical codes.

At the same time, dimension $12$ appears to be a natural candidate for further progress. Recent work revealed an unexpected richness in the space of optimal known configurations: there exist at least $1579$ pairwise non-isometric kissing arrangements of size $840$ in $\mathbb R^{12}$~\cite{takhanov2026classificationindependentsetssigned}. This abundance of extremal structures suggests that the geometry of the problem is far from rigid and raises the possibility that many more optimal configurations remain undiscovered.

Perhaps more importantly, the family of $840$-point arrangements discovered in this paper reveals a previously unknown continuum of extremal configurations in $\mathbb{R}^{12}$, demonstrating the existence of degrees of freedom that had not been observed before.
The geometric structure of these deformations, together with the associated Clebsch-based $48$-point subsystems, suggested a highly non-random initialization for logarithmic Riesz energy minimization. Starting from such structurally informed initial data, we were able to obtain a spherical code of size $841$ in $\mathbb R^{12}$ with maximal inner product strictly smaller than $\frac{1}{2}$. The corresponding Gram matrix is provided in the accompanying \href{https://github.com/k-nic/841_in_12D}{github} repository.

\section{The construction of the $48$-system}
Most of the kissing arrangements of size $840$ in $\mathbb R^{12}$ discovered so far share a common underlying structure. They consist of $60$ vectors supported on the first factor $\mathbb R^6\times\{\mathbf 0\}$, another $60$ vectors supported on the second factor $\{\mathbf 0\}\times\mathbb R^6$, and a collection of $720$ additional vectors, which we call \emph{bridge vectors}. The first two blocks may be viewed as independent six-dimensional configurations, while the bridge vectors encode the interaction between the two copies of $\mathbb R^6$.

\begin{wrapfigure}{r}{0.25\textwidth}
\centering
\includegraphics[width=0.2\textwidth]{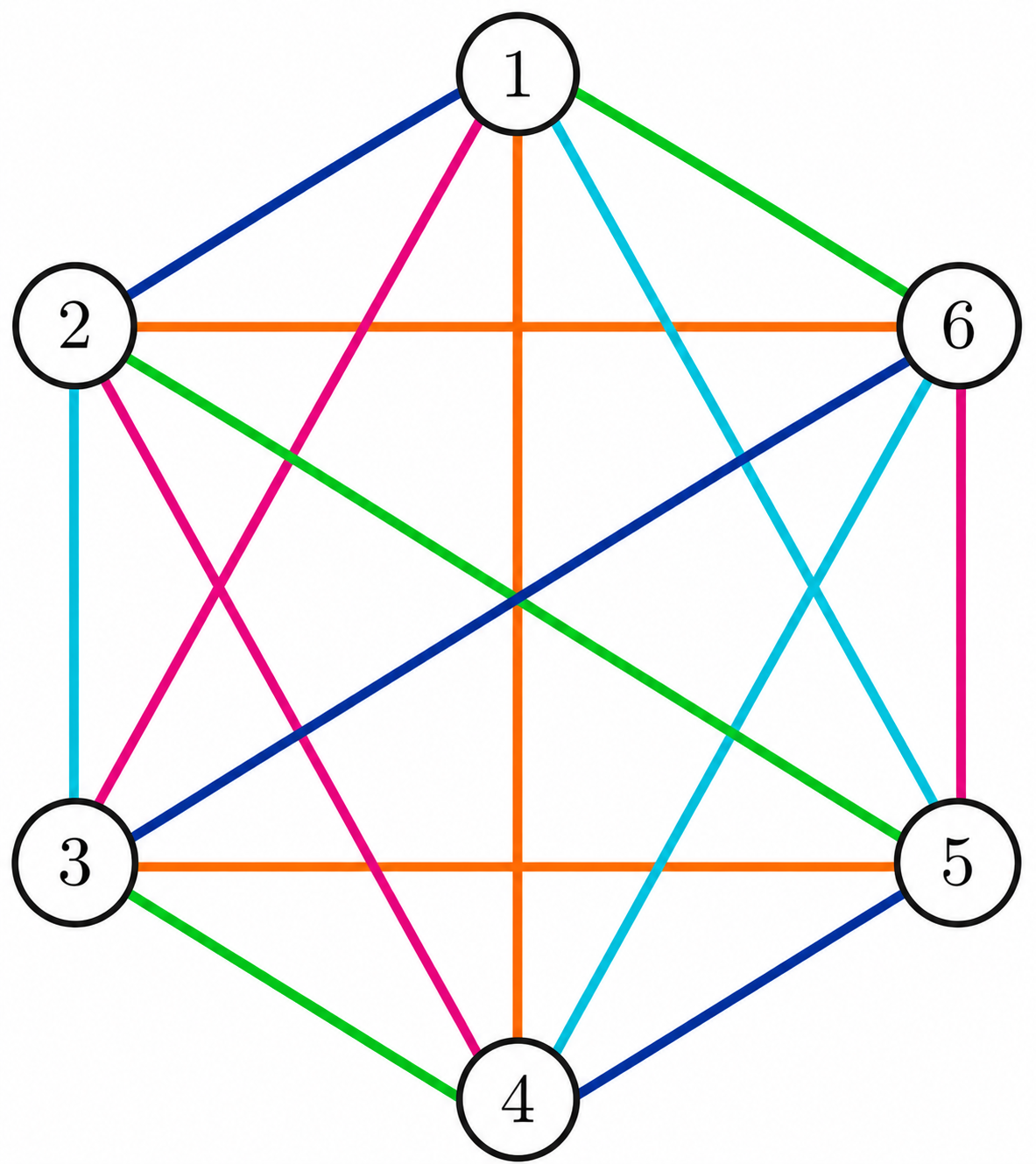}
\caption{\small 1-factorization of $K_6$.}\label{1-factor}
\end{wrapfigure}
The bridge construction is based on the unique $1$-factorization of the complete graph $K_6$ (see Figure~\ref{1-factor}). Recall that the edges of $K_6$ can be partitioned into five perfect matchings. Given edges $\{i,j\}$ and $\{k,\ell\}$ of the same color in $K_6$, we form vectors in $\mathbb R^{12}$ supported on the four coordinates
\[
i,\ j,\ 6+k,\ 6+\ell .
\]
On these four coordinates we put arbitrary signs, each with absolute value $\frac{1}{2}$, and put zeros elsewhere. Thus each bridge vector has the form
\[
v
=
\frac12\left(
\varepsilon_i e_i+\varepsilon_j e_j
+\varepsilon_k e_{6+k}+\varepsilon_\ell e_{6+\ell}
\right),
\qquad
\varepsilon_i,\varepsilon_j,\varepsilon_k,\varepsilon_\ell\in\{\pm1\}.
\]
There are $15$ choices for the edge $\{i,j\}$, $3$ choices for the edge $\{k,\ell\}$ (given $\{i,j\}$), and $2^4$ possible choices of signs. Hence this construction gives
\[
15\cdot 3\cdot 2^4=720
\]
bridge vectors. Together with the two $60$-point six-dimensional blocks, these vectors form a kissing arrangement of size $840$ in $\mathbb R^{12}$.

\subsection{Motivation behind the  $48$-system: a numerical evidence}
Suppose that the set of vectors $\{v_1,...,v_{60}\}\subseteq {\mathbb R}^6$ is such that the vectors $\{v_1,...,v_{60}\}\times \{{\mathbf 0}\}\subseteq {\mathbb R}^6$, together with $720$ bridge vectors, form a kissing arrangement in ${\mathbb R}^{12}$. Thus, we have the following constraints on the coordinates of vectors:
\begin{itemize}
\item $v_i^\top v_i =1, i\in [60]$;
\item $v_i^\top v_j \leq \frac{1}{2}$ for any distinct indices $i,j\in [60]$ (mutual interaction);
\item $|v_{k}|+|v_{l}|\leq 1$ for any distinct indices $k,l\in [6]$ and any $v\in \{v_1,...,v_{60}\}$ (interaction with bridge vectors).
\end{itemize}
We searched for such systems numerically using the Adam optimizer. Using free variables $x_i\in {\mathbb R}^6$, $i=1,...,60$,
for normalized vectors $v_i=\frac{x_i}{\|x_i\|}$, we minimized the objective
\[
\frac1T
\log\!\left(
\sum_{1\le i<j\le 60}
\exp\!\left(
T\max\!\left(v_i^\top v_j-\frac12,0\right)
\right)
+
\sum_{i=1}^{60}
\sum_{1\le k<l\le 6}
\sum_{\varepsilon,\delta\in\{\pm1\}}
\exp\!\left(
\frac{T}{2}
\max\!\left(
\varepsilon v_{ik}
+\delta v_{il}
-1,0
\right)
\right)
\right),
\]
with \(T=100\).

All systems of vectors that we were able to find numerically contained  signed coordinate vectors, i.e. $\{\pm e_i\mid i\in [6]\}$. So, further we assumed that $\{v_i\}_{i=49}^{60}$ are signed coordinate vectors. This led us to study the remaining $48$ vectors. Since the remaining vectors should have an angle at least $60$ degrees with signed coordinate vectors, the third inequality on coordinates  simplifies to
\begin{itemize}
\item $|v_{k}|\leq \frac{1}{2}$ for any $k\in [6]$ and any $v\in \{v_1,...,v_{48}\}$.
\end{itemize}
Let us denote the set of all systems $\{v_1,...,v_{48}\}$ that satisfy the latter inequalities by $\mathcal{S}_{48}$, and call every element in  $\mathcal{S}_{48}$ {\em a $48$-system}. The resulting dataset of such configurations is available in the accompanying github repository as a zip-archive saved\_48\_systems.zip.
\begin{figure}[htbp]
\begin{center}
\begin{tikzpicture}[
    scale=.85,
    every node/.style={font=\small},
    point/.style={circle,fill=black,inner sep=1.45pt},
    hidden/.style={densely dashed,gray!70,line width=0.55pt},
    visible/.style={black,line width=0.85pt},
    axis/.style={-{Latex[length=2.3mm]},line width=0.65pt}
]

\def\R{3.0}
\def\eqrx{3.0}
\def\eqry{0.72}
\def\ly{1.55}
\def\my{-1.55}
\def\lrx{2.57}
\def\lry{0.62}

\draw[black,line width=0.9pt] (0,0) circle (\R);

\draw[axis] (3.75,-2.35) -- (3.75,2.55);
\draw[thin] (3.62,\my) -- (3.88,\my) node[right=3pt] {$x_1=-\frac12$};
\draw[thin] (3.62,0) -- (3.88,0) node[right=3pt] {$x_1=0$};
\draw[thin] (3.62,\ly) -- (3.88,\ly) node[right=3pt] {$x_1=\frac12$};

\draw[hidden]  (0,\my) ellipse[x radius=\lrx,y radius=\lry];
\draw[visible] (\lrx,\my) arc[start angle=0,end angle=-180,x radius=\lrx,y radius=\lry];

\draw[hidden]  (0,0) ellipse[x radius=\eqrx,y radius=\eqry];
\draw[visible] (\eqrx,0) arc[start angle=0,end angle=-180,x radius=\eqrx,y radius=\eqry];

\draw[hidden]  (0,\ly) ellipse[x radius=\lrx,y radius=\lry];
\draw[visible] (\lrx,\ly) arc[start angle=0,end angle=-180,x radius=\lrx,y radius=\lry];

\foreach \i/\ang in {1/225,2/270,3/315,7/20,11/55,15/110,16/150}{
    \coordinate (X\i) at ({\eqrx*cos(\ang)},{\eqry*sin(\ang)});
    \coordinate (B\i) at ({\lrx*cos(\ang)},{\ly+\lry*sin(\ang)});
    \coordinate (C\i) at ({\lrx*cos(\ang)},{\my+\lry*sin(\ang)});
}

\tikzset{liftarrow/.style={-{Latex[length=2.0mm]},gray!65,line width=0.45pt,densely dashed}}
\foreach \i in {1,2,3,15,16}{
    \draw[liftarrow] (X\i) -- (B\i);
    \draw[liftarrow] (X\i) -- (C\i);
}

\foreach \i in {1,2,3,7,11,15,16}{
    \node[point] at (X\i) {};
    \node[point] at (B\i) {};
    \node[point] at (C\i) {};
}

\node[below left,font=\scriptsize]  at (X1) {$x_1$};
\node[below,font=\scriptsize]       at (X2) {$x_2$};
\node[below,font=\scriptsize]       at (X3) {$x_3$};
\node[below,font=\scriptsize]       at (X15) {$x_{15}$};
\node[below right,font=\scriptsize] at (X16) {$x_{16}$};

\node[above left,font=\scriptsize]  at (B1) {$b_1$};
\node[above,font=\scriptsize]       at (B2) {$b_2$};
\node[above,font=\scriptsize]       at (B3) {$b_3$};
\node[above,font=\scriptsize]       at (B15) {$b_{15}$};
\node[above right,font=\scriptsize] at (B16) {$b_{16}$};

\end{tikzpicture}
\caption{Schematic view of the $48$-system in $\mathbb{R}^6$: the equatorial layer $x_1=0$ and the two lifted layers $x_1=\pm\frac12$.
}
\end{center}
\end{figure}

Surprisingly, the $48$-systems that we were able to find using numerical modeling (and all explicit examples that we are aware of), all consist of three layers:
\begin{enumerate}
\item The equator vectors, that is $16$ vectors
\[
\{0\}\times \{x_1, \cdots, x_{16}\}.
\]
\item The $+\frac{1}{2}$ floor vectors, i.e. 
\[
\left\{\frac{1}{2}\right\}\times \{b_1, \cdots, b_{16}\}.
\]
\item The $-\frac{1}{2}$ floor vectors, i.e. 
\[
\left\{-\frac{1}{2}\right\}\times \{b_1, \cdots, b_{16}\},
\]
where vectors $\{b_i\}$ are the same for  $+\frac{1}{2}$ and $-\frac{1}{2}$ floors.
\end{enumerate}
A key observation from numerically obtained examples is that the set of pairs $(x_i,x_j)$ with an obtuse angle between them, form the Clebsch graph. Moreover, elements $\{b_i\}$ can be arranged in such a way that the contact graph whose set of vertices is $[16]$ and the set of edges is
$$
i\sim j \quad\Longleftrightarrow\quad
    x_i^\top b_j = \frac{1}{2},
$$
is the same Clebsch graph. 

Recall that the Clebsch graph is 5-regular, so for a fixed $i$, $x_i$ has 5 neighbours, denoted $x_{j_1}$, ..., $x_{j_5}$. For the examples obtained numerically, we made the most natural assumption that $b_i$ can be recovered as a linear combination of the neighbours of $x_i$, that is 
$$
b_i = \sum_{k=1}^5 \beta_k x_{j_k}.
$$
If the neighbours are linearly independent, this can be done uniquely, that is
\begin{equation}\label{recover}
[\beta_k]_{k=1}^5 = \frac{1}{2}\big(\left[x_{j_k}^\top x_{j_l}\right]_{k,l=1}^{5}\big)^{-1}[1,1,1,1,1]^\top,
\end{equation}
so the $\frac{1}{2}$-floor system $\{b_1, \cdots, b_{16}\}$ can be recovered from $\{x_1, \cdots, x_{16}\}$ analytically. It turns out that this natural assumption explains the majority of the numerical examples we observed, and consequently we focused mainly on this case. An example of a $48$-system that violates the latter condition is given below  for completeness. 

\begin{remark}[A $48$-system with degenerate Clebsch equator]
Note that the $48$-point subsystem of the classical kissing arrangement of Leech and Sloane also possesses an equator set whose graph of obtuse pairs is the Clebsch graph. However, in this case a set of neighbours of a given vector is linearly dependent.  
Consider the following set of $48$ vectors normalized by $2$:
\[
\begin{array}{cccccc}
x_1 & x_2 & x_3 & x_4 & x_5 & x_6\\
\hline
\pm1 & \pm1 & \pm1 & \pm1 & 0 & 0\\
\pm1 & \pm1 & 0 & 0 & \pm1 & \pm1\\
0 & 0 & \pm1 & \pm1 & \pm1 & \pm1
\end{array}
\]
The equator with respect to the first coordinate is
\[
 E
=
\left\{
\left(0,0,\frac{\varepsilon_3}{2},
\frac{\varepsilon_4}{2},
\frac{\varepsilon_5}{2},
\frac{\varepsilon_6}{2}\right)
\mid
\varepsilon_i\in\{\pm 1\}
\right\}.
\]
Thus, the equator consists of the $16$ vertices of a four-dimensional cube.

The obtuseness graph on $E$ is equivalent to the graph whose vertices are the sign vectors
$\varepsilon=(\varepsilon_3,\varepsilon_4,\varepsilon_5,\varepsilon_6)
\in\{-1,+1\}^4$
and where two distinct vertices $\varepsilon,\delta$ are adjacent whenever
the corresponding vectors have negative inner product.  This inner product is negative if and only if the Hamming distance between $\varepsilon$ and $\delta$ is $3$ or $4$.  This graph is
isomorphic to the Clebsch graph (an exact definition of it can be found in the beginning of the next section).

However, all vectors in $E$
lie in the coordinate subspace
$\{x\in\mathbb R^6\mid x_1=x_2=0\}$,
and hence span only a $4$-dimensional subspace.  
\end{remark}
\subsection{Equator description}
We use the following definition of the canonical Clebsch graph. Let
\[
    \Omega=\left\{
    \varepsilon\in\{+1,-1\}^5\mid
    \varepsilon_1\varepsilon_2\varepsilon_3\varepsilon_4\varepsilon_5=1
    \right\}.
\]
Thus $|\Omega|=16$. 
For $\varepsilon,\delta\in\Omega$, define
\[
    (\varepsilon,\delta)\in E_c
    \quad\Longleftrightarrow\quad
   d_{\rm H}(\varepsilon,\delta)=4,
\]
where $ d_{\rm H}(\varepsilon,\delta) = |\{r\mid\varepsilon_r\neq\delta_r\}|$ is the Hamming distance. Note that $d_{\rm H}(\varepsilon,\delta)\in \{0,2,4\}$.

The latter definition of edges is equivalent to the obtuseness relation between vectors. Indeed, we have
$$
\varepsilon^\top \delta = 5-2d_{\rm H}(\varepsilon,\delta)<0 \Leftrightarrow d_{\rm H}(\varepsilon,\delta)=4.
$$
Any graph isomorphic to $(\Omega,E_c)$ is called the Clebsch graph. 

Let $G=([16],E)$ be the Clebsch graph, and let
$N(i)$ denote the set of neighbours of $i$.

Let a collection of vectors
$\{x_1,\ldots,x_{16}\}\subseteq S^4\subset\mathbb R^5$
be given. Let us formulate main conditions that this collection should satisfy in order to define a $48$-system.
Let
\[
\Gamma=\left[x_i^\top x_j\right]_{i,j=1}^{16}
\]
be the Gram matrix of the system.
For $A,B\subseteq [16]$, let
$\Gamma[A,B]$
denote the submatrix of $\Gamma$ whose rows are indexed by $A$
and whose columns are indexed by $B$.
For every $i\in[16]$, define
$\Gamma_i=\Gamma[N(i),N(i)]$.

Elements of the equator set, after adding signed coordinate vectors $\pm e_i$, should become a kissing arrangement, therefore, we require:
\begin{enumerate}
\item $|x_{ik}|\leq \frac{1}{2}$ for all $i\in[16]$ and $k\in[5]$;

\item $x_i^\top x_j\leq \frac{1}{2}$ for all $i\neq j$;
\end{enumerate}
Also, pairs of vectors with an obtuse angle between them form a Clebsch graph, i.e.
\begin{enumerate}
\setcounter{enumi}{2}
\item $x_i^\top x_j<0$ if and only if $(i,j)\in E$;
\end{enumerate}
We also need a number of conditions, which are equivalent to requiring that: (a) analytically recovered system $\{b_1, ...,b_{16}\}$  lies on the sphere $\frac{\sqrt{3}}{2}{\mathbb S}^{5}$; (b) $\left\{\frac{1}{2}\right\}\times \{b_1, \cdots, b_{16}\}$ forms a kissing arrangement in ${\mathbb R}^6$; (c) components of $b_i$ are in $[-\frac{1}{2},\frac{1}{2}]$. That is, we have
\begin{enumerate}
\setcounter{enumi}{3}
\item for every $i\in[16]$, the vectors
$\{x_j\mid j\in N(i)\}$ are linearly independent and
$\mathbf 1^\top \Gamma_i^{-1}\mathbf 1 =3$, where ${\mathbf 1}=[1,1,1,1,1]^\top$;

\item for every $i\in[16]$ and every coordinate
$r\in[5]$,
\[
\left|
\sum_{k\in N(i)}
(\Gamma_i^{-1}\mathbf 1)_k\,x_{kr}
\right|
\leq 1;
\]

\item for every $i,t\in[16]$ with
$t\notin N(i)$,
\[
\mathbf 1^\top
\Gamma_i^{-1}
\Gamma[N(i),\{t\}]
\leq 1;
\]

\item for every distinct $i,j\in[16]$,
\[
\mathbf 1^\top
\Gamma_i^{-1}
\Gamma[N(i),N(j)]
\Gamma_j^{-1}
\mathbf 1
\leq 1.
\]
\end{enumerate}
The set of systems $\{x_1,...,x_{16}\}$ satisfying conditions 1-7 is called {\em the equator set} and is denoted by $\mathcal{E}_{16}$.
Note that the latter set of inequalities and identities can all be written as polynomial constraints on components of $x_i$'s.

Let us formulate theorem that describes how $48$-system can be constructed from the equator vectors. 
\begin{theorem}
Let $\{x_1,...,x_{16}\}\in \mathcal{E}_{16}$. 
Then there are $48$ vectors in $\mathbb R^6$ of the form
\[
    (0,x_i),\qquad
    \left(\frac12,b_i\right),\qquad
    \left(-\frac12,b_i\right),
    \qquad i=1,\ldots,16,
\]
that constitute a kissing arrangement. Moreover all coordinates of all these
vectors have absolute value at most $1/2$, that is this system belongs to $\mathcal{S}_{48}$.
\end{theorem}

\begin{proof}
For each $i\in[16]$, let
\[
    X_i=\begin{bmatrix}x_j\end{bmatrix}_{j\in N(i)}
\]
be the $5\times 5$ matrix whose columns are the vectors $x_j$,
$j\in N(i)$, in the fixed order used to define
$\Gamma_i=\Gamma[N(i),N(i)]$. By definition,
$\Gamma_i=X_i^\top X_i$. 
Define
\[
    b_i=\frac{1}{2} X_i\Gamma_i^{-1}\mathbf 1,
\]
which is the matrix form of the equation~\eqref{recover}.
First we compute the norm of $b_i$. Since
$\mathbf 1^\top \Gamma_i^{-1}\mathbf 1=3$, we get
\[
\begin{aligned}
    \|b_i\|^2
    &=
    \frac14
    \mathbf 1^\top
    \Gamma_i^{-1}
    X_i^\top X_i
    \Gamma_i^{-1}
    \mathbf 1       \\
    &=
    \frac14
    \mathbf 1^\top
    \Gamma_i^{-1}
    \Gamma_i
    \Gamma_i^{-1}
    \mathbf 1       =
    \frac14
    \mathbf 1^\top
    \Gamma_i^{-1}
    \mathbf 1
    =
    \frac34,
\end{aligned}
\]
that is $b_i\in \frac{\sqrt{3}}{2}{\mathbb S}^4$.
Hence
\[
    \left\|\left(\pm\frac12,b_i\right)\right\|^2
    =
    \frac14+\|b_i\|^2
    =
    1.
\]
Also, $\|(0,x_i)\|=\|x_i\|=1$. Thus, all $48$ vectors $(0,x_i)$,$\left(\pm\frac12,b_i\right)$ lie on $S^5$.

Next, we check the inner products. For two equator vectors,
\[
    (0,x_i)^\top(0,x_j)=x_i^\top x_j\leq \frac12
    \qquad (i\neq j),
\]
by the definition of $\mathcal E_{16}$ (condition 2).

Now consider one floor vector and one equator vector. If $j\in N(i)$, then
\[
    X_i^\top b_i
    =
    \frac12 X_i^\top X_i\Gamma_i^{-1}\mathbf 1
    =
    \frac12\Gamma_i\Gamma_i^{-1}\mathbf 1
    =
    \frac12\mathbf 1,
\]
and therefore
\[
    b_i^\top x_j=\frac12.
\]
If $j\notin N(i)$, then by the defining inequalities of
$\mathcal E_{16}$ (condition 6),
\[
    b_i^\top x_j
    =
    \frac12
    \mathbf 1^\top
    \Gamma_i^{-1}
    \Gamma[N(i),\{j\}]
    \leq
    \frac12.
\]
Thus, for all $i,j$,
\[
    \left(\pm\frac12,b_i\right)^\top(0,x_j)
    =
    b_i^\top x_j
    \leq
    \frac12.
\]

Now consider two vectors in the same floor. For $i\neq j$, using the
definition of $b_i$ and $b_j$, we have
\[
\begin{aligned}
    b_i^\top b_j
    &=
    \frac14
    \mathbf 1^\top
    \Gamma_i^{-1}
    X_i^\top X_j
    \Gamma_j^{-1}
    \mathbf 1  \\
    &=
    \frac14
    \mathbf 1^\top
    \Gamma_i^{-1}
    \Gamma[N(i),N(j)]
    \Gamma_j^{-1}
    \mathbf 1
    \leq
    \frac14,
\end{aligned}
\]
again by the definition of $\mathcal E_{16}$ (condition 7). Hence
\[
    \left(\frac12,b_i\right)^\top
    \left(\frac12,b_j\right)
    =
    \frac14+b_i^\top b_j
    \leq
    \frac12,
\]
and similarly
\[
    \left(-\frac12,b_i\right)^\top
    \left(-\frac12,b_j\right)
    =
    \frac14+b_i^\top b_j
    \leq
    \frac12.
\]

Finally, for vectors in opposite floors and $i, j\in [16]$,
\[
    \left(\frac12,b_i\right)^\top
    \left(-\frac12,b_j\right)
    =
    -\frac14+b_i^\top b_j.
\]
Since $\|b_i\|=\|b_j\|=\sqrt3/2$, we have
\[
    b_i^\top b_j\leq \|b_i\|\|b_j\|=\frac34.
\]
Therefore
\[
    -\frac14+b_i^\top b_j
    \leq
    -\frac14+\frac34
    =
    \frac12.
\]

Thus every pair of distinct vectors among the $48$ constructed vectors
has inner product at most $1/2$. Hence they form a kissing arrangement
in $\mathbb R^6$.

It remains to check the coordinate bound. For the equator vectors this is
part of the definition of $\mathcal E_{16}$ (condition 1):
\[
    |x_{ik}|\leq \frac12.
\]
For the floor vectors, the first coordinate is $\pm 1/2$, and the
remaining coordinates are the coordinates of $b_i$. The defining
coordinate constraints of $\mathcal E_{16}$ (condition 5) give
\[
    |(b_i)_k|\leq \frac12
    \qquad\text{for all }i\in[16],\ k\in[5].
\]
Therefore, all coordinates of all $48$ vectors have absolute value at most
$1/2$.
\end{proof}
The following is a key example of an element from $\mathcal{E}_{16}$.
\begin{example}\label{48-clebsch} Vertices of the canonical Clebsch graph cannot occupy the equator due to the violation of the Condition 4, but their skewed version can.
Let
\[
    q_1=q_2=q_3=q_4=\frac29,
    \qquad
    q_5=\frac19.
\]
Index the $16$ vertices by even sign vectors $\varepsilon=(\varepsilon_1,\ldots,\varepsilon_5)\in\{+1,-1\}^5$ such that
    $ \varepsilon_1\cdots\varepsilon_5=1$.
Define
\[
    x_\varepsilon
    =
    \left(
        \frac{\sqrt2}{3}\varepsilon_1,
        \frac{\sqrt2}{3}\varepsilon_2,
        \frac{\sqrt2}{3}\varepsilon_3,
        \frac{\sqrt2}{3}\varepsilon_4,
        \frac13\varepsilon_5
    \right).
\]
Let us show that $\{x_\varepsilon\}$ belongs to $\mathcal E_{16}$.
Indeed,
$\|x_\varepsilon\|^2
    =
    4\cdot\frac29+\frac19=1$,
and all coordinates have absolute value at most $\frac{1}{2}$ (condition 1).

If two distinct even sign vectors differ in a set $D\subseteq[5]$, then
$|D|=2$ or $|D|=4$, and
\[
    x_\varepsilon^\top x_\delta
    =
    1-2\sum_{r\in D}q_r.
\]
If $|D|=2$, then this inner product is either
$1-2\left(\frac29+\frac29\right)=\frac19$
or
$1-2\left(\frac29+\frac19\right)=\frac13$.
Thus, it is positive. If $|D|=4$, then it is either
$1-2\cdot\frac89=-\frac79$
or
$1-2\cdot\frac79=-\frac59$.
Thus, it is negative exactly for Clebsch edges, so the condition 3 is satisfied. In particular,
$x_\varepsilon^\top x_\delta\leq \frac{1}{2}$ for all distinct
$\varepsilon,\delta$ (condition 2).

It remains to check conditions for the lifted vectors $b_i$. Fix a vertex $\varepsilon$. Its five
neighbours are obtained by 
flipping each of five coordinates of $-\varepsilon$. Therefore, the Gram matrix of the
neighbours is, up to simultaneous row and column permutation,
\[
    \Gamma_i=
    \begin{pmatrix}
    1 & 1/9 & 1/9 & 1/9 & 1/3\\
    1/9 & 1 & 1/9 & 1/9 & 1/3\\
    1/9 & 1/9 & 1 & 1/9 & 1/3\\
    1/9 & 1/9 & 1/9 & 1 & 1/3\\
    1/3 & 1/3 & 1/3 & 1/3 & 1
    \end{pmatrix}.
\]
A direct computation gives
\[
    \Gamma_i^{-1}\mathbf 1=\begin{pmatrix}
\frac{9}{8} & 0 & 0 & 0 & -\frac{3}{8}\\
0 & \frac{9}{8} & 0 & 0 & -\frac{3}{8}\\
0 & 0 & \frac{9}{8} & 0 & -\frac{3}{8}\\
0 & 0 & 0 & \frac{9}{8} & -\frac{3}{8}\\
-\frac{3}{8} & -\frac{3}{8} & -\frac{3}{8} & -\frac{3}{8} & \frac{3}{2}
\end{pmatrix}\mathbf 1
    =
    \left(\frac34,\frac34,\frac34,\frac34,0\right)^\top,
\]
and hence
\[
    \mathbf 1^\top\Gamma_i^{-1}\mathbf 1
    =
    4\cdot\frac34=3.
\]
Thus the condition 4 holds for every vertex.

It can be easily checked that neighbours of $x_\varepsilon$ are independent, so vectors $b_\varepsilon$ are analytically defined. Moreover, for this system the corresponding vectors $b_i$ are given by
\[
    b_\varepsilon
    =
    -\left(
        \frac{\sqrt2}{4}\varepsilon_1,
        \frac{\sqrt2}{4}\varepsilon_2,
        \frac{\sqrt2}{4}\varepsilon_3,
        \frac{\sqrt2}{4}\varepsilon_4,
        \frac12\varepsilon_5
    \right)\in \frac{\sqrt{3}}{2}{\mathbb S}^4.
\]
Hence
\[
    |(b_\varepsilon)_r|\leq \frac12
\]
for every coordinate $r$ (condition 5).

Finally, one checks directly that
\[
    b_\varepsilon^\top x_\delta = -\frac{1}{6}\sum_{k=1}^5\varepsilon_k \delta_k  = -\frac{1}{6}(5-2d_{\rm H}(\varepsilon,\delta))\leq \frac12
\]
for all $\varepsilon,\delta\in \Omega$ (condition 6), with equality exactly when the two sign
vectors differ in four coordinates, and  
\[
    b_\varepsilon^\top b_\delta= \frac{1}{8}\sum_{k=1}^4\varepsilon_k\delta_k+\frac{1}{4}\varepsilon_5\delta_5\leq \frac{3}{4}-2(\frac{1}{8}+\frac{1}{8})=\frac14
    \qquad
    \text{for }\varepsilon\neq\delta\text{ (condition 7)}.
\]
Therefore, all defining inequalities of $\mathcal E_{16}$ are satisfied.
\end{example}
We observed numerically that $48$-systems are supported on some continuous manifold rather than on a discrete set. Indeed, the following theorem is in full correspondence with that numerical result.
\begin{theorem}
Let $A\in O(4)$ and put
$Q_A=
\begin{pmatrix}
A&0\\
0&1
\end{pmatrix}
\in O(5)$. 
Define
$x_\varepsilon^A=Q_Ax_\varepsilon$ where $\{x_\varepsilon\}\in\mathcal E_{16}$ is the equator set from the previous example.
If
\[
\max_{1\le r\le 4}\sum_{s=1}^4 |a_{rs}| 
\le
\frac{3}{2\sqrt2},
\]
then
$\{x_\varepsilon^A\mid \varepsilon\in\Omega\}\in\mathcal E_{16}$.
\end{theorem}
Note that $\|A\|_\infty=
\max_{1\le r\le 4}\sum_{s=1}^4 |a_{rs}|$, and for orthogonal matrices close to $I$ the latter condition is satisfied due to $\|I\|_\infty=1$.
\begin{proof}
Since $Q_A$ is orthogonal, it preserves all inner products. Therefore the
Gram matrix of the system $\{x_\varepsilon^A\}$ is the same as the Gram
matrix of $\{x_\varepsilon\}$. Hence all conditions in the definition of
$\mathcal E_{16}$ that depend only on inner products are unchanged.

It remains only to check the conditions 1 and 5. 
For $r=1,\ldots,4$, we have
\[
(x_\varepsilon^A)_r
=
\frac{\sqrt2}{3}
\sum_{s=1}^4 a_{rs}\varepsilon_s.
\]
Thus
\[
|(x_\varepsilon^A)_r|
\le
\frac{\sqrt2}{3}
\sum_{s=1}^4 |a_{rs}|.
\]
By assumption,
\[
\frac{\sqrt2}{3}
\sum_{s=1}^4 |a_{rs}|
\le
\frac{\sqrt2}{3}\cdot\frac{3}{2\sqrt2}
=
\frac12.
\]
Therefore
$|(x_\varepsilon^A)_r|\leq \frac12$
 for $r=1,\ldots,4$.
For the fifth coordinate,
$(x_\varepsilon^A)_5=\frac13\varepsilon_5$,
so
$|(x_\varepsilon^A)_5|=\frac13\leq\frac12$. 
Hence all coordinates of all vectors $x_\varepsilon^A$ have absolute value
at most $1/2$.

It remains to check condition 5. The lifted vectors $b_\varepsilon^A$
are defined by 
\[
b_\varepsilon^A=Q_A b_\varepsilon,
\]
where $b_\varepsilon
=
-\left(
\frac{\sqrt2}{4}\varepsilon_1,
\frac{\sqrt2}{4}\varepsilon_2,
\frac{\sqrt2}{4}\varepsilon_3,
\frac{\sqrt2}{4}\varepsilon_4,
\frac12\varepsilon_5
\right)$,
because the local Gram matrices are unchanged and the neighbour coordinate matrices
are multiplied by $Q_A$.

For $r=1,\ldots,4$,
\[
|(b_\varepsilon^A)_r|
=
\frac{\sqrt2}{4}
\left|
\sum_{s=1}^4 a_{rs}\varepsilon_s
\right|
\le
\frac{\sqrt2}{4}
\sum_{s=1}^4 |a_{rs}|.
\]
Since
\[
\sum_{s=1}^4 |a_{rs}|
\le
\frac{3}{2\sqrt2},
\]
we get
\[
|(b_\varepsilon^A)_r|
\le
\frac{\sqrt2}{4}\cdot\frac{3}{2\sqrt2}
=
\frac38
<
\frac12.
\]
For the fifth coordinate,
$|(b_\varepsilon^A)_5|=\frac12$. 
Therefore
$|(b_\varepsilon^A)_r|\leq \frac12$
for all $r=1,\ldots,5$, which is exactly condition 5.

Therefore all defining conditions of $\mathcal E_{16}$ are satisfied, and
so $\{x_\varepsilon^A\mid \varepsilon\in\Omega\}\in\mathcal E_{16}$.
\end{proof}
\section{Computational search.}
The computational search was based on a continuation scheme for logarithmic Riesz energies~\cite{cohn2026variationsfivedimensionalspherepackings}. We started from the $840$-point kissing arrangement in $\mathbb R^{12}$ consisting of two copies of the $60$-point Clebsch configuration (that is, vectors of the $48$-system described in example~\ref{48-clebsch} and $12$ signed coordinate vectors) occupying ${\mathbb R}^6\times \{{\mathbf 0}\}$ and $\{{\mathbf 0}\}\times  {\mathbb R}^6$ together with the $720$ bridge vectors arising from the $K_6$ one-factorization construction. To obtain a configuration of size $841$, an additional vector was appended and initialized as
\[
x_{841}
=
\frac1{\sqrt{12}}
(\varepsilon_1,\ldots,\varepsilon_{12}),
\qquad
\varepsilon_i\in\{-1,1\},
\]
where the signs $\varepsilon_i$ were chosen independently and uniformly at random.

The resulting $841$-point configuration
\[
X=\{x_1,\ldots,x_{841}\}\subset S^{11}
\]
was optimized by minimizing the logarithmic Riesz energy
\[
E_s(X)
=
\log\!\left(
\sum_{1\le i<j\le 841}
\|x_i-x_j\|^{-s}
\right),
\]
using the Adam optimizer in $64$ independent batches. Following a standard continuation strategy, the exponent $s$ was gradually increased through a prescribed schedule, allowing the optimization to focus progressively on the pairs with the largest inner products. Independent runs were performed with different random choices of the additional sign vector. The best configurations were subsequently polished by further optimization procedures.

The source code used for the search and optimization, together with the Gram matrix of the best configuration found, is available at \href{https://github.com/k-nic/841_in_12D}{github}.

The final Gram matrix defines a set of $841$ unit vectors in $\mathbb R^{12}$ with maximal pairwise inner product
\[
\max_{i\ne j}\langle x_i,x_j\rangle
=
0.499999937751
<
\frac12.
\]
Hence it yields a kissing arrangement of size $841$ in dimension $12$.

\subsection{Beyond $841$ vectors}
We believe that the initialization scheme introduced in this paper has the potential
to produce kissing arrangements with even larger cardinalities, provided that
sufficient optimization resources are devoted to the search.

Indeed, we have already obtained a spherical code of size $842$ in
$\mathbb S^{11}$ with maximal pairwise inner product
$0.500901018602$.
Although we have not yet succeeded in polishing this configuration into a kissing
arrangement, it lies very close to the kissing threshold. Similarly, we found a
spherical code of size $844$ in $\mathbb S^{11}$ with maximal pairwise inner
product
$0.5044042665$.
The corresponding Gram matrices are available in the github repository
accompanying this paper.
\bibliographystyle{IEEEtran}
\bibliography{lit}

\end{document}